\documentclass[aps,prl,twocolumn,superscriptaddress]{revtex4-1}
\usepackage[letterpaper, margin=1.8cm]{geometry}
\usepackage{graphicx}
\usepackage{dcolumn}
\usepackage{bm}
\usepackage{amsmath}
\usepackage[english]{babel}
\usepackage[utf8]{inputenc}
\usepackage[T1]{fontenc}
\usepackage{verbatim}
\usepackage{physics}

\begin{document}
\title{A four-qubit germanium quantum processor }
\author{N.W.~Hendrickx}
\email{n.w.hendrickx@tudelft.nl}
\affiliation{QuTech and Kavli Institute of Nanoscience, Delft University of Technology, Lorentzweg 1, 2628 CJ Delft, The Netherlands}
\author{W.I.L.~Lawrie}
\author{M.~Russ}
\author{F.~van~Riggelen}
\author{S.L.~de~Snoo}
\author{R.N.~Schouten}
\affiliation{QuTech and Kavli Institute of Nanoscience, Delft University of Technology, Lorentzweg 1, 2628 CJ Delft, The Netherlands}
\author{A.~Sammak}
\affiliation{QuTech and Netherlands Organisation for Applied Scientific Research (TNO), Stieltjesweg 1, 2628 CK Delft, The Netherlands}
\author{G.~Scappucci}
\affiliation{QuTech and Kavli Institute of Nanoscience, Delft University of Technology, Lorentzweg 1, 2628 CJ Delft, The Netherlands}
\author{M.~Veldhorst}
\email{m.veldhorst@tudelft.nl}
\affiliation{QuTech and Kavli Institute of Nanoscience, Delft University of Technology, Lorentzweg 1, 2628 CJ Delft, The Netherlands}
	
\begin{abstract}
The prospect of building quantum circuits using advanced semiconductor manufacturing positions quantum dots as an attractive platform for quantum information processing. Extensive studies on various materials have led to demonstrations of two-qubit logic in gallium arsenide, silicon, and germanium. However, interconnecting larger numbers of qubits in semiconductor devices has remained an outstanding challenge. Here, we demonstrate a four-qubit quantum processor based on hole spins in germanium quantum dots. Furthermore, we define the quantum dots in a two-by-two array and obtain controllable coupling along both directions. Qubit logic is implemented all-electrically and the exchange interaction can be pulsed to freely program one-qubit, two-qubit, three-qubit, and four-qubit operations, resulting in a compact and high-connectivity circuit. We execute a quantum logic circuit that generates a four-qubit Greenberger-Horne-Zeilinger state and we obtain coherent evolution by incorporating dynamical decoupling. These results are an important step towards quantum error correction and quantum simulation with quantum dots.
\end{abstract}
\maketitle

Fault-tolerant quantum computers utilizing quantum error correction \cite{terhal_quantum_2015} to solve relevant problems \cite{reiher_elucidating_2017} will rely on the integration of millions of qubits. Solid-state implementations of physical qubits have intrinsic advantages to accomplish this formidable challenge and remarkable progress has been made using qubits based on superconducting circuits \cite{arute_quantum_2019}. While the development of quantum dot qubits has been at a more fundamental stage, their resemblance to the transistors that constitute the building block of virtually all our electronic hardware promises excellent scalability to realize large-scale quantum circuits \cite{loss_quantum_1998, vandersypen_interfacing_2017}. Fundamental concepts for quantum information, such as the coherent rotation of individual spins \cite{koppens_driven_2006} and the coherent coupling of spins residing in neighboring quantum dots \cite{petta_coherent_2005}, were first implemented in gallium arsenide heterostructures. The low disorder in the quantum well allowed the construction of larger arrays of quantum dots and to realize two-qubit logic using two singlet-triplet qubits \cite{shulman_demonstration_2012}. However, spin qubits in group III-V semiconductors suffer from hyperfine interactions with nuclear spins that severely limit their quantum coherence. Group IV materials naturally contain higher concentrations of isotopes with a net-zero nuclear spin and can furthermore be isotopically enriched \cite{itoh_isotope_2014} to contain only these isotopes. In silicon electron spin qubits, quantum coherence can therefore be sustained for a long time \cite{muhonen_storing_2014, veldhorst_addressable_2014} and single qubit logic can be implemented with fidelities exceeding 99.9 \% \cite{yoneda_quantum-dot_2018, yang_silicon_2019}. By exploiting the exchange interaction between two spin qubits in adjoining quantum dots or closely separated donor spins, two-qubit logic could be demonstrated \cite{veldhorst_two-qubit_2015, zajac_resonantly_2018, watson_programmable_2018, huang_fidelity_2019, he_two-qubit_2019, madzik_conditional_2020, petit_universal_2020}. Silicon, however, suffers from a large effective mass and valley degeneracy \cite{zwanenburg_silicon_2013}, which has hampered progress beyond two-qubit demonstrations.

Holes in germanium are emerging as a promising alternative \cite{scappucci_germanium_2020} that combine favorable properties such as zero nuclear spin isotopes for long quantum coherence \cite{itoh_high_1993}, low effective mass and absence of valley states \cite{lodari_light_2019} for relaxed requirements on device design, low charge noise for a quiet qubit environment \cite{lodari_low_2020}, and low disorder for reproducible quantum dots \cite{hendrickx_gate-controlled_2018, lawrie_quantum_2020}. In addition, strained germanium quantum wells defined on silicon substrates are compatible with semiconductor manufacturing \cite{pillarisetty_academic_2011}. Furthermore, hole states can exhibit strong spin-orbit coupling that allows for all-electric operation \cite{maurand_cmos_2016, watzinger_germanium_2018, hendrickx_fast_2020} and that removes the need for microscopic components such as microwave striplines or nanomagnets, which is particularly beneficial for the fabrication and operation of two-dimensional qubit arrays. The realization of strained germanium quantum wells in undoped heterostructures \cite{sammak_shallow_2019} has led to remarkable progress. In two year's time, germanium has progressed from the formation of stable quantum dots and quantum dot arrays \cite{hendrickx_gate-controlled_2018, lawrie_quantum_2020, van_riggelen_two-dimensional_2020}, to demonstrations of single qubit logic \cite{hendrickx_single-hole_2020}, long spin lifetimes \cite{lawrie_spin_2020}, and the realization of fast two-qubit logic in germanium double quantum dots \cite{hendrickx_fast_2020}. 

\begin{figure*}[htp]
\includegraphics{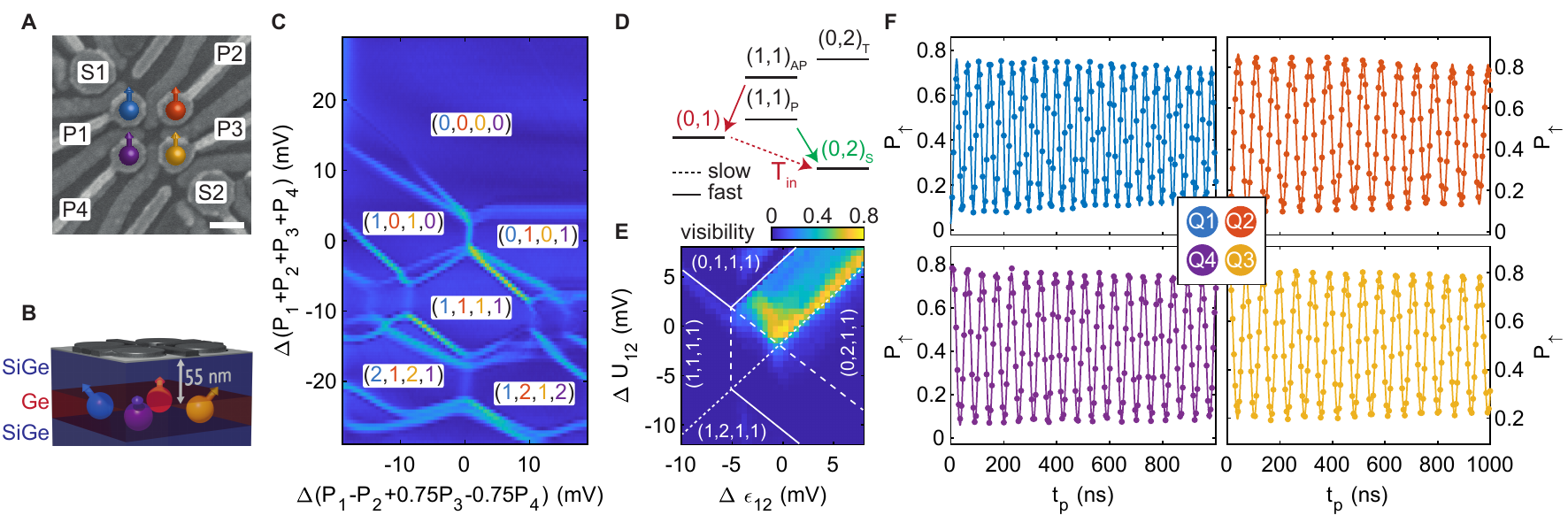}
\caption{\textbf{Four germanium hole spin qubits.}
(\textbf{A}) Scanning electron microscope image of the four quantum dot device. We define qubits underneath the four plunger gates indicated by P1-P4. The qubits can be measured using the two charge sensors S1 and S2. The scale bar corresponds to 100~nm.
(\textbf{B}) Schematic drawing of the Ge/SiGe heterostructure. Starting from a silicon wafer, a germanium quantum well is grown in between two Si$_{0.2}$Ge$_{0.8}$ layers at a depth of 55~nm from the semiconductor/dielectric interface.
(\textbf{C}) Four quantum dot charge stability diagram as a function of two virtual gates. At the vertical and diagonal bright lines a hole can tunnel between two quantum dots or a quantum dot and its reservoir respectively. As a result of the virtual axes, the addition lines of the different quantum dots have different slopes, allowing for an easy distinction of the different charge occupations indicated in the white boxes as (Q1, Q2, Q3, Q4).
(\textbf{D}) Energy diagram illustrating the latched Pauli spin blockade readout. When pulsing from the (1,1) charge state to the (0,2) charge state, only the polarized triplet states allow the holes to move into the same quantum dot, leaving an (0,2) charge state (green). Interdot tunneling is blocked for the two antiparallel spin states and as a result the hole on the first quantum dot will subsequently tunnel to the reservoir leaving an (0,1) charge state (red), locking the different spin states into different charge states.
(\textbf{E}) Readout visibility as defined by the difference in readout between either applying no rotation and a $\pi$-rotation to Q2. The readout point is moved around the (1,1)-(0,2) anticrossing of the Q1Q2 system and a clear readout window can be observed bounded by the different (extended) reservoir transition lines indicated by the dotted lines.
(\textbf{F}) The qubits can be rotated by applying a microwave tone resonant with the Zeeman splitting of the qubit. Coherent Rabi rotations can be observed as a function of the microwave pulse length $t_\text{p}$ for all qubits Q1-Q4.
}
\label{fig:fig1}
\end{figure*}

\begin{figure*}[htp]
\includegraphics{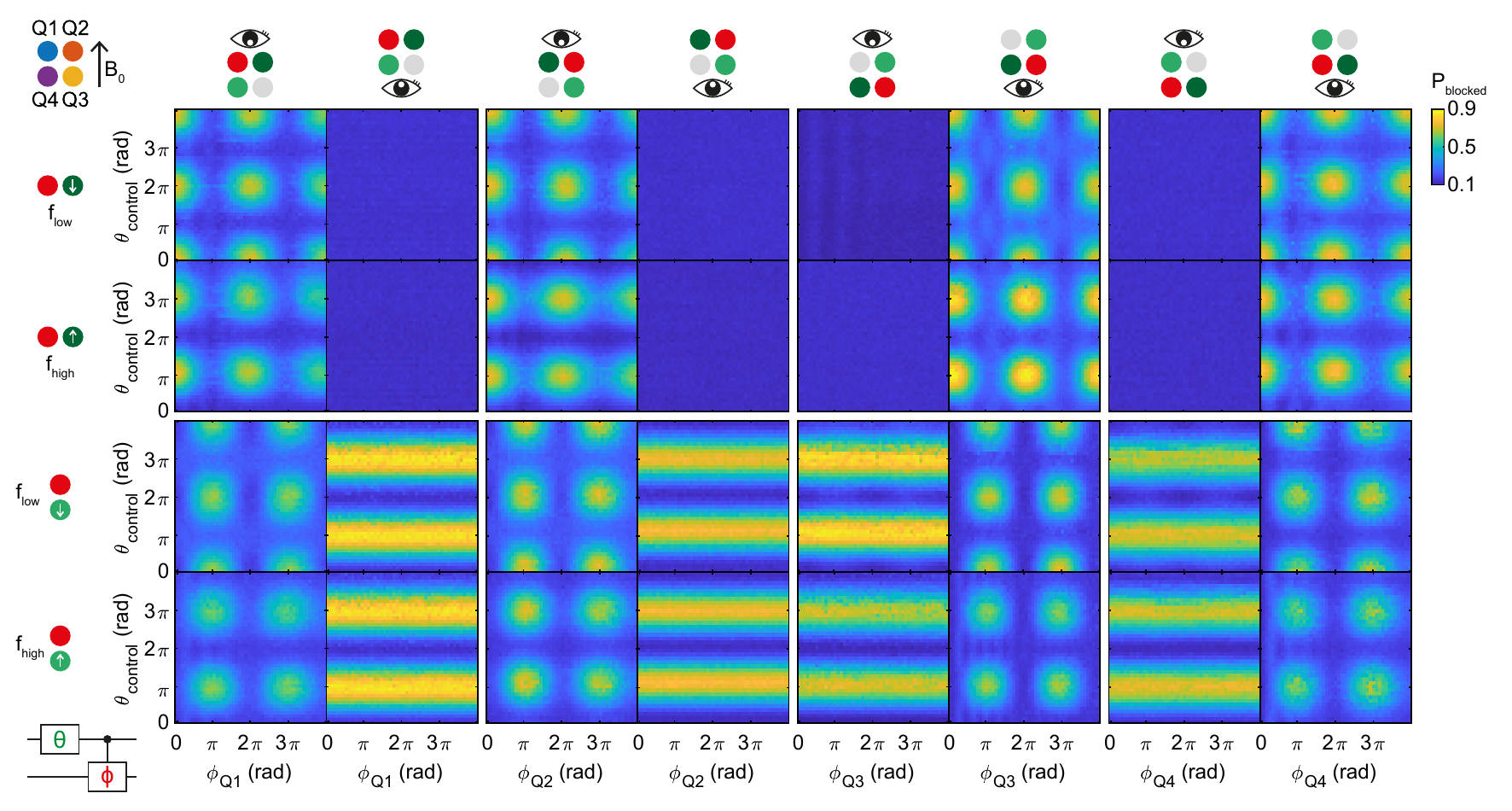}
\caption{\textbf{Controlled rotations between all nearest-neighbor qubit pairs.} 
By selectively enabling the exchange interaction between each pair of qubits, we can implement two-qubit controlled rotations (CROTs). The pulse sequence consists of a single preparation gate with length $\theta$ on the control qubit (labeled green), followed by a controlled rotation on one of the resonance lines of the target qubit (labelled in red). Both qubit pairs Q1Q2 and Q3Q4 are read out in single-shot mode and the position of the eye on top of each column indicates the respective readout pair. Each of the four main columns corresponds to conditional rotations on a different qubit as indicated by the red dot. Rows one and two show the results for the horizontal interaction (dark green), while rows three and four show the two-qubit interaction for the vertical direction (light green) with respect to the external magnetic field, as indicated in the top left. Rows one and three correspond to driving the lower frequency $f_\text{low}$ conditional resonance line, while rows two and four show driving of the other resonance line $f_\text{high}$.
}
\label{fig:fig2}
\end{figure*}

\begin{figure*}[htp]
\includegraphics{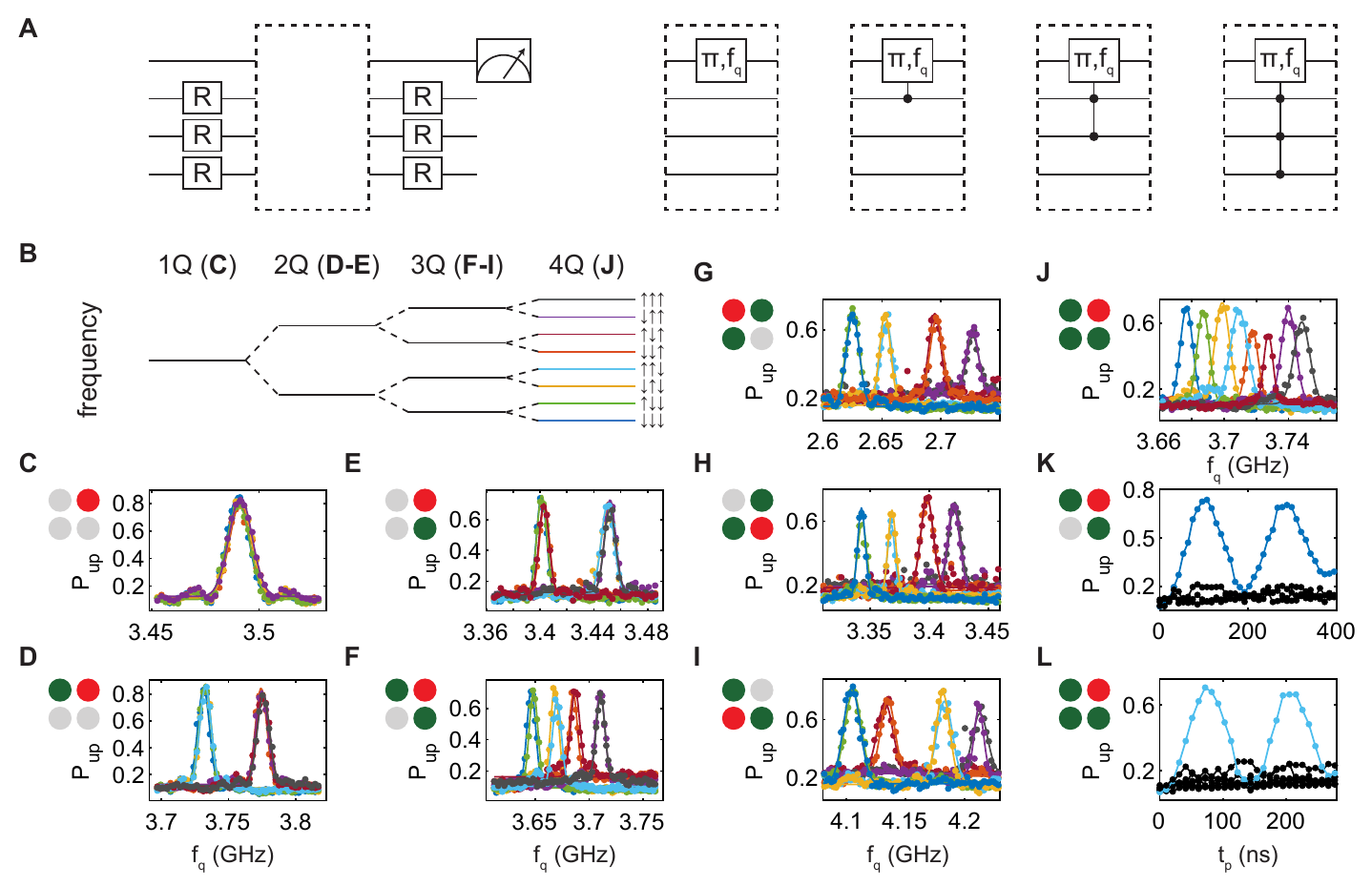}
\caption{\textbf{Resonant one, two, three, and four-qubit gates.} 
    (\textbf{A}) Circuit diagram of the experiment performed in panels \textbf{C-L}. All eight permutations of the three control qubit eigenstates are prepared, with R being either no pulse or a $\pi$-pulse on the respective qubit. Next, the resonance frequency of the target qubit is probed using a $\pi$-rotation with varying frequency $f_q$. Finally, the prepared qubits are projected back and the target qubit state is measured. By changing the different interdot couplings $J$, we can switch between resonant single, two, three, and four-qubit gates as indicated in the dashed boxes.
    (\textbf{B}) Turning on the exchange interaction between the different qubit pairs splits the resonance frequency in two, four, and eight for 1, 2 and 3 enabled pairs respectively. The colors of the line segments correspond to the colors in panels \textbf{C-L}.
    (\textbf{C}) By turning all exchange interactions off, the qubit resonance frequency of Q2 is independent of the prepared state of the other three qubits, resulting in an effective single-qubit rotation.
    (\textbf{D-E}) By turning on a single exchange interaction $J_{12}$ (\textbf{D}) or $J_{23}$ (\textbf{E}), the resonance line splits in two.
    (\textbf{F-I}), Turning on both exchange interactions to the neighboring quantum dots results in the resonance line splitting in four, for Q2 (\textbf{F}), Q1 (\textbf{G}), Q3 (\textbf{H}), Q4 (\textbf{I}) respectively.
    (\textbf{J}) Turning on the exchange interactions between three pairs of quantum dots $J_{12}$, $J_{23}$, $J_{41}$ splits the resonance line in eight.
    (\textbf{K-L}) Resonant driving of the three-qubit gate (\textbf{K}) and the four-qubit gate (\textbf{L}) with Q2 being the target qubit, shows Rabi driving as a function of pulse length $t_\text{p}$, demonstrating the coherent evolution of the operation.
    }
\label{fig:fig3}
\end{figure*}

\begin{figure*}[htp]
\includegraphics{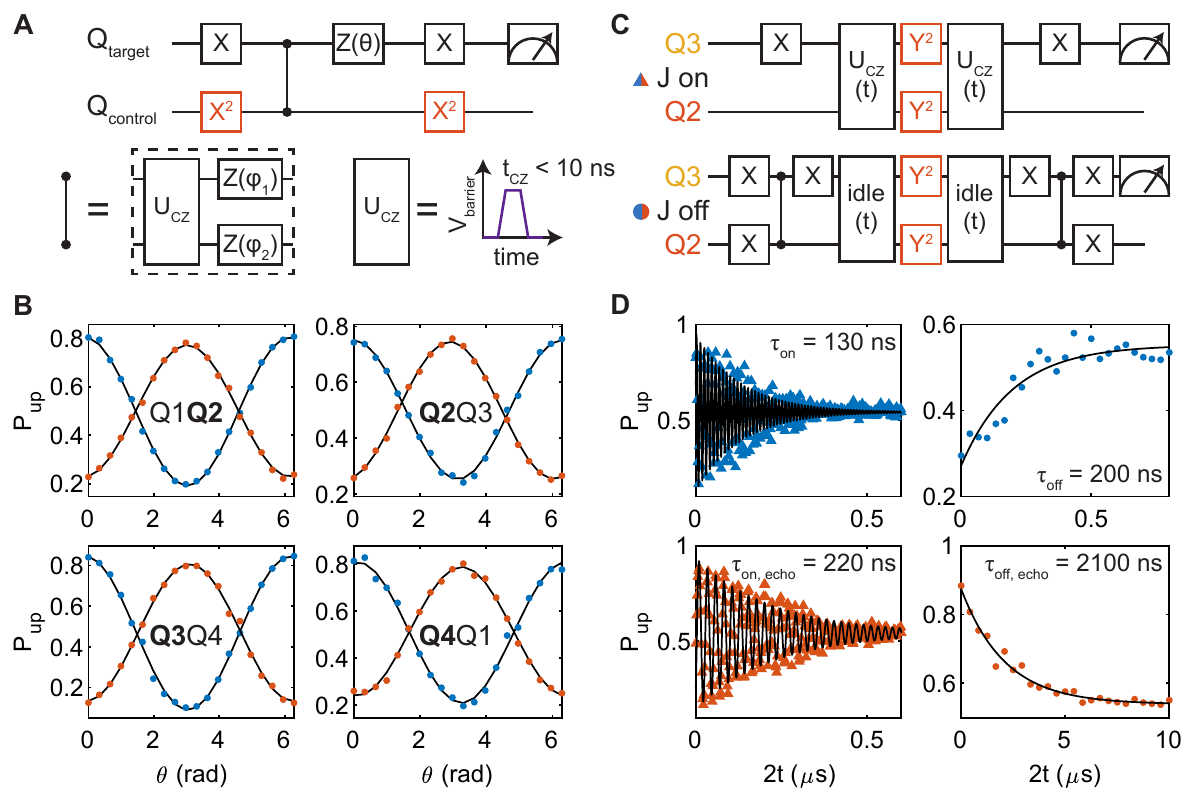}
\caption{\textbf{Controlled phase gate and dynamical decoupling.} 
    (\textbf{A}) Circuit diagram of the experiment performed in panel \textbf{B}. The controlled phase gate is probed by performing a Ramsey sequence on the target qubit for both basis states of the control qubit. The phase of the second $\pi/2$ (X) gate is swept by performing an update of the microwave phase through quadrature modulation. Additionally, a phase update is performed on both the target and control qubit to compensate for any single qubit phases picked up as a result of the gate pulsing to achieve a controlled-Z (CZ) gate.
    (\textbf{B}) The spin-up probability of the target qubit (in bold) as a function of the phase $\theta$ of the second X gate for the control qubit initialized in the $\ket{\downarrow}$ (blue) and $\ket{\uparrow}$ (red) state. Measurements for the inverted target and control qubits in Fig.~S10. By applying an exchange pulse and single qubit phase updates, we achieve a CZ gate at $\theta = 0$~rad.
    (\textbf{C}) Circuit diagrams of the experiment performed in panel \textbf{D}. The phase coherence throughout the two-qubit experiment is probed using a Ramsey sequence, both for the case with $J$ on (top) and off (bottom) and both with (orange) and without (blue) applying an echo pulse.
    (\textbf{D}) Spin-up probability as a function of the experiment length, for the situation with exchange on (left, triangles) and off (right, circles). From the decay data we extract characteristic decay times $\tau$ of $\tau_\text{on} = 130$~ns, $\tau_\text{on, echo} = 220$~ns, $\tau_\text{off} = 200$~ns, and $\tau_\text{off, echo} = 2100$~ns (details in Materials and Methods).
    }
\label{fig:fig4}
\end{figure*}

 \begin{figure*}[htp]
\includegraphics{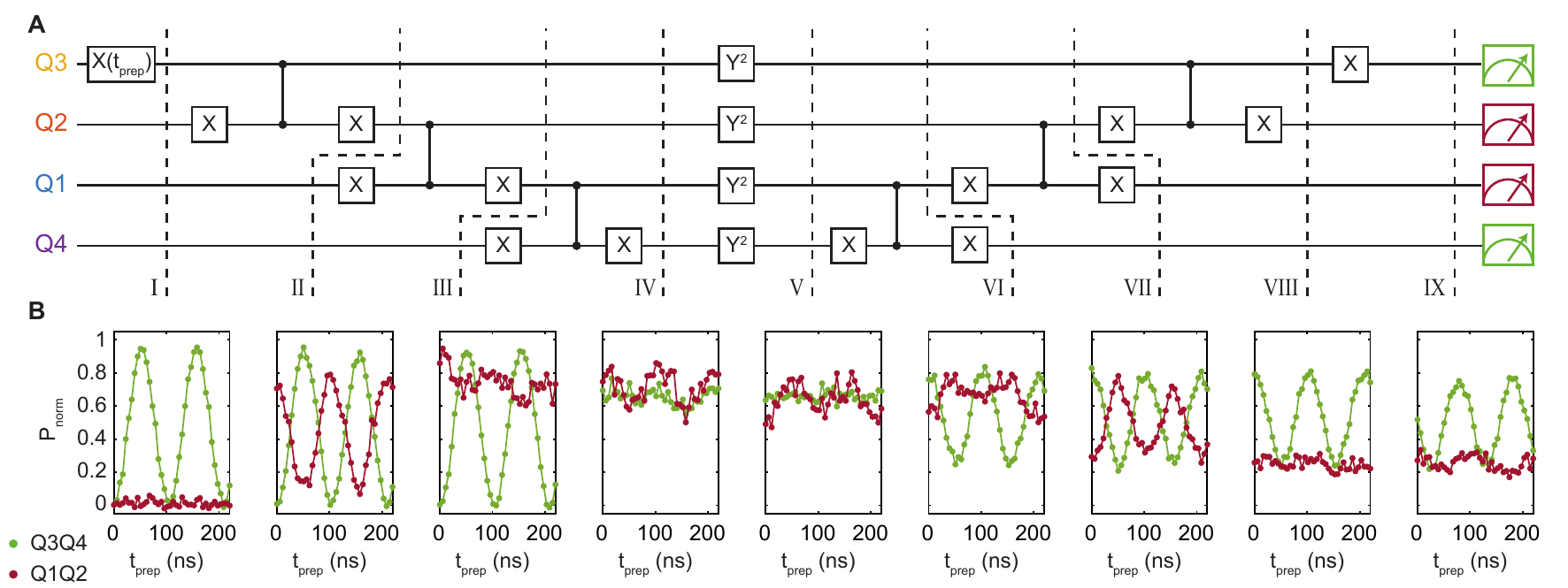}
\caption{\textbf{Coherent generation of a four-qubit Greenberger-Horne-Zeilinger (GHZ) state.} 
    (\textbf{A-B}) A four-qubit GHZ state is created by applying three sequential two-qubit gates, each consisting of an X-CZ-X gate circuit. Next, a Y$^2$ decoupling pulse is applied, after which we disentangle the GHZ state again (circuit diagram in \textbf{A}). Pulses pictured in the same column are applied simultaneously. The initial state of Q3 is varied by applying a preparation rotation of length $t$. For different stages throughout the algorithm (dashed lines), we measure the non-blocked state probability as a function of $t$ for both the Q1Q2 and Q3Q4 readout system, normalized to their respective readout visibility. At the end of the algorithm the qubit states correspond to the initial single qubit rotation, and the clear oscillations confirm the coherent evolution of the algorithm from isolated qubit states to a four-qubit GHZ state. 
    (\textbf{B}).
    }
\label{fig:fig5}
\end{figure*}

Here, we advance semiconductor quantum dots beyond two-qubit realizations and execute a four-qubit quantum circuit using a two-dimensional array of quantum dots. We achieve this by defining the four-qubit system on the spin states of holes in gate-defined germanium quantum dots. Fig.~\ref{fig:fig1}A shows a scanning-electron-microscopy (SEM) image of the germanium quantum processor. The quantum dots are defined in a strained germanium quantum well on a silicon substrate (Fig.~\ref{fig:fig1}B) \cite{lodari_low_2020} using a double layer of electrostatic gates and contacted by aluminum ohmic contacts. A negative potential on plunger gates P1-P4 accumulates a hole quantum dot underneath that serves as qubit Q1-Q4, which can be coupled to neighboring quantum dots through dedicated barrier gates. In addition, two quantum dots are placed to the side of the two-by-two array, and the total system comprises six quantum dots. Via an external tank circuit, we configure these additional two quantum dots as radio frequency (rf) charge sensors for rapid charge detection. Using the combined signal of both charge sensors \cite{van_riggelen_two-dimensional_2020}, we measure the four quantum dot stability diagram as shown in Fig.~\ref{fig:fig1}C. Making use of two virtual gate axes, we arrange the reservoir addition lines of the four quantum dots to have different relative slopes of approximately $-1, +1, -0.75, 0.75$~mV/mV for Q1, Q2, Q3, and Q4 respectively. Well defined charge regions (indicated as (Q1,Q2,Q3,Q4) in the white boxes) are observed, with vertical anticrossings marking the different interdot transitions. The high level of symmetry in the plot is a sign of comparable gate lever arms and quantum dot charging energies, confirming the uniformity in this platform and simplifying the operation of quantum dot arrays.

For the qubit readout we make use of Pauli-spin blockade to convert the spin states into a charge signal that can be detected by the sensors. In germanium, however, the spin-orbit coupling can significantly lower the spin lifetime during the readout process, in particular when the spin-orbit field is perpendicular to the external magnetic field, reducing the readout fidelity \cite{danon_pauli_2009, hendrickx_single-hole_2020}. Here, we overcome this effect by making use of a latched readout process \cite{harvey-collard_high-fidelity_2018}. During the readout process, as illustrated in Fig.~\ref{fig:fig1}D, a hole can tunnel spin-selectively to the reservoir as a result of different tunnel rates of both quantum dots to the reservoir. After this process, the system is locked in this charge state for the slow reservoir tunnel time $T_\text{in}$. We achieve this effect by pulsing into the area in the (0,2) charge region bounded by the extended (1,1)-(0,1) (fast) and the extended (1,1)-(1,2) (slow) transitions (dotted lines in Fig.~\ref{fig:fig1}E). When the interdot tunneling into the (0,2) charge state is blocked, the hole in the first quantum dot will quickly tunnel into the reservoir. This locks the spin state in the metastable (0,1) charge state, with the decay to the (0,2) ground state governed by the slow tunnel rate $T_\text{in}$ between the second quantum dot and the reservoir. 
The high level of control in germanium allows the tuning of $T_\text{in}$ to arbitrarily long time scales by changing the potential applied to the corresponding reservoir barrier gates. We set $T_\text{in, Q2} = 200~\mu$s and $T_\text{in, Q4} = 2.4$~ms (Fig.~S2), both significantly longer than the signal integration time $T_\text{int} = 10~\mu$s. We operate in a parity readout mode where we observe both antiparallel spin states to be blocked (Fig.~S3AB). We speculate this is caused by the strong spin-orbit coupling mixing the parallel (1,1) states with the (0,2) state, and causing strong relaxation of the upper parallel spin state. By both increasing the interdot coupling and elongating the ramp between the manipulation and readout point, we can transition into a state selective readout where only the $\ket{\downarrow\uparrow}$ state results in spin blockade (Fig.~S3CD), with a slightly reduced readout visibility.

In our experiments, we configure the system such that the spin-orbit field is oriented along the direction of the external magnetic field $B_0 = 1.05$~T. This minimizes relaxation and we project all qubit measurements onto this readout direction, thus reading out qubit pairs Q1Q2 and Q3Q4. Each charge sensor can detect transitions in both qubit pairs, but is mostly sensitive to their respective nearby quantum dots. We maximize the readout visibility as defined by the difference between the readout of a spin-up and spin-down state by scanning the readout level around the relevant anticrossing. This is illustrated for the Q1Q2 pair in Fig.~\ref{fig:fig1}E, where a clear readout window with maximum visibility can be observed bounded between the (extended) reservoir transitions of the two quantum dots. 

Coherent rotations can be implemented by applying electric microwave signals to the plunger gates that define the qubits, exploiting the spin-orbit coupling for fast driving \cite{watzinger_germanium_2018, bulaev_electric_2007}. We initialize the system in the $\ket{\downarrow\downarrow\downarrow\downarrow}$ state by sequentially pulsing both the Q1Q2 and Q3Q4 double quantum dot systems from their respective $(0,2)_\text{S}$ states adiabatically into their $(1,1)_{\text{T}_-}$ states. We then perform the qubit manipulations, after which we perform the spin readout as described above. We observe qubit resonances at $f_\text{Q1}=2.304$~GHz, $f_\text{Q2}=3.529$~GHz, $f_\text{Q3}=3.520$~GHz, and $f_\text{Q4}=3.882$~GHz, corresponding to effective $g$-factors of $g_\text{Q1}=0.16$, $g_\text{Q2}=0.24$, $g_\text{Q3}=0.24$, and $g_\text{Q4}=0.26$. We note that these $g$-factors can be electrically modulated using nearby gates as a means to ensure individual qubit addressability. Fig.~\ref{fig:fig1}F shows the single-shot spin-up probability $P_\uparrow$ for each of the four qubits after applying an on-resonant microwave burst with increasing time duration $t_\text{p}$, resulting in coherent Rabi oscillations. 

To quantify the quality of the single qubit gates, we perform benchmarking of the Clifford group \cite{knill_randomized_2008} (Fig.~S4) and find single qubit gate fidelities exceeding 99 \% for all qubits. The fidelity of Q3 is even above 99.9 \%, thereby comparing to benchmarks for quantum dot qubits in isotopically purified silicon \cite{yoneda_quantum-dot_2018, yang_silicon_2019}. We find spin lifetimes between $T_1=1-16$~ms (Fig.~S5), comparable to values reported before for holes in planar germanium \cite{lawrie_spin_2020}. Furthermore, we observe $T_2^*$ to be between 150-400~ns for the different qubits (Fig.~S6A), but are able to extend phase coherence up to $T_2^\text{CPMG}=100~\mu$s by performing Carr-Purcell-Meiboom-Gill (CPMG) refocusing pulses (Fig.~S6C), more than two orders of magnitude larger than previously reported for hole quantum dot qubits \cite{maurand_cmos_2016, watzinger_germanium_2018, hendrickx_fast_2020}. This indicates the qubit phase coherence is mostly limited by low-frequency noise, which is confirmed by the predominantly $1/f^\alpha$ noise spectrum we observe by Ramsey and dynamical decoupling noise spectroscopy (Fig.~S7). This noise could originate in the nuclear spin bath present in germanium, which could be mitigated by isotopic enrichment. Alternatively, it could be caused by charge noise acting on the spin state through the spin-orbit coupling and it is predicted that the sensitivity to this type of noise could be mitigated by careful optimization of the electric field environment \cite{wang_suppressing_2019} or moving to a multi-hole charge occupancy, screening the influence of charge impurities \cite{barnes_screening_2011}, potentially enabling even higher fidelity operations.

Universal quantum logic can be accomplished by combining the single qubit rotations with a two-qubit entangling gate. We implement this using a conditional rotation (CROT) gate, where the resonance frequency of the target qubit depends on the state of the control qubit, mediated by the exchange interaction $J$ between the two quantum dots. The exchange interaction between the quantum dots is controlled using a virtual barrier gate (details in Materials and Methods), coupling the two quantum dots while keeping the detuning and on-site energy of the quantum dots constant and close to the charge-symmetry point. We demonstrate CROT gates between all four pairs of quantum dots in Fig.~\ref{fig:fig2}, proving that spin qubits can be coupled in two dimensions. A sequence of qubit pulses is applied, as indicated in the diagram, consisting of a single qubit control pulse (green) and a target qubit two-qubit pulse (red). We vary the length of both the control pulse $\theta_\text{control}$ as well as the length of the target qubit pulse $\phi_\text{Q1-Q4}$, with $t_\text{p}(\phi=\pi)=50-110$~ns (details in Table S1). The conditional rotations are performed on all four target qubits (main four columns) for both the horizontally interacting qubits (rows 1 and 2), as well as the vertically interacting qubits (rows 3 and 4), by driving the $\ket{\downarrow\downarrow}$-$\ket{\uparrow\downarrow}$ transitions with $f_\text{low}$ (rows 1 and 3), as well as the inverse $\ket{\downarrow\uparrow}$-$\ket{\uparrow\uparrow}$ transitions with $f_\text{high}$ (rows 2 and 4), with $\ket{\text{Q}_\text{target}\text{Q}_\text{control}}$. We then perform a measurement on both readout pairs by sequentially pulsing the Q1Q2 (left sub-columns), and the Q3Q4 qubit pairs (right sub-columns) to their respective readout points. Because the target qubit resonance frequency depends on the control qubit state, the conditional rotation is characterized by the fading in and out of the target qubit rotations as a function of the control qubit pulse length. The pattern is therefore shifted by a $\pi$ rotation on the control qubit, for driving the two separate transitions. When driving the $\ket{\downarrow\downarrow}$-$\ket{\uparrow\downarrow}$ transition of the qubit pairs used for readout (row 1), we apply an additional single-qubit $\pi$-pulse to the preparation qubit for symmetry, since the control qubit also serves as the readout ancilla. When the control qubit is in a different readout pair as the target qubit (rows 3 and 4), we can independently observe the single qubit control, and two-qubit target qubit rotations in the two readout systems. By setting the pulse length equal to $\phi_\text{Q}=\pi$, a fast CX gate can be obtained within approximately $t_p=100$~ns between all of the four qubit pairs.

To demonstrate full control over the coupling between the different qubits, we measure the qubit resonance frequency as a function of the eight possible permutations of the different basis states of the other three qubits, as illustrated in Fig.~\ref{fig:fig3}A-B. Without any exchange present, the resonance frequency of the target qubit should be independent on the preparation of the other three qubits, as schematically depicted in Fig.~\ref{fig:fig3}C. When the exchange to one of the neighboring quantum dots is enabled by pulsing the virtual barrier gate, the resonance line splits in two, allowing for the operation of the CROT gate, as is shown for both the Q1-Q2 and Q2-Q3 interactions in Fig.~\ref{fig:fig3}D and E respectively. When both barriers to the nearest-neighbors are pulsed open at the same time, we observe a fourfold splitting of the resonance line (Fig.~\ref{fig:fig3}F-I). This allows the performance of a resonant $i$-Toffoli three-qubit gate (Fig.~\ref{fig:fig3}K and Fig.~S8), which has theoretically been proposed as an efficient manner to create the Toffoli, Deutsch, and Fredkin gates \cite{gullans_protocol_2019}. We observe a difference in the efficiency at which the different conditional rotations can be driven, as can also be seen from the width of the resonance peaks in Fig.~\ref{fig:fig3}F-I. This is expected to happen when the exchange energy is comparable to the difference in Zeeman splitting and is caused by the mixing of the basis states due to the exchange interaction between the holes~\cite{hetenyi_exchange_2020} (details in Materials and Methods). Finally, we open three of the four virtual barriers and observe the resonance line splitting in eight, being different for all eight permutations of the control-qubit preparation states (Fig.~\ref{fig:fig3}J). This enables us to execute a resonant four-qubit gate and in Fig.~\ref{fig:fig3}L we show the coherent operation of a three-fold conditional rotation (see Fig.~S8 for the coherent operation of the other resonance lines). 

While the demonstration of these conditional rotations can be beneficial for the simulation of larger coupled spin systems, the ability to dynamically control the exchange interaction allows for faster two-qubit operations \cite{veldhorst_two-qubit_2015, watson_programmable_2018}. We efficiently implement controlled phase (CPHASE) gates between the different qubit pairs by adiabatically pulsing the exchange interaction using the respective virtual barrier gate. Increasing the exchange interaction, the antiparallel spin states will shift in energy with respect to the parallel spin states, giving rise to a conditional phase accumulation. We control the length and size of the voltage pulse (Fig S9) to acquire a CZ gate, in which the antiparallel spin states accumulate a phase of exactly $\pi$ with respect to the parallel spin states. We demonstrate this in Fig.~\ref{fig:fig4}A-B, where we employ a Ramsey sequence to measure the conditional phase. After the exchange pulse $U_\text{CZ}$ we apply a software Z gate to both the target and control qubits to compensate for individual single qubit phases. As a result of the large range over which the exchange interaction can be controlled, we achieve fast CZ gates that are executed well within 10~ns for all qubit pairs (details in Table S2).

To prepare our system for quantum algorithms, we implement decoupling pulses into the multi-qubit sequences to extend phase coherence, as demonstrated in Fig.~\ref{fig:fig4}C-D. To probe the effect of a decoupling pulse when exchange is on (Fig.~\ref{fig:fig4}D, left, triangles), we perform a CPHASE gate between qubits Q2 and Q3 and compare the decay of the resulting exchange oscillations as a function of the operation time for the situations with (orange) and without (blue) a Y$^2$ echo pulse. We observe an extended duration for the conditional phase rotations of $\tau=220$~ns when applying a decoupling pulse, compared to $\tau=~130$~ns for a standard CPHASE gate. A more relevant situation however, is the coherence of the two-qubit entangled state. We probe this by entangling Q2 and Q3 by forming the $\ket{\Psi^+}$ Bell state and letting the system evolve for time $2t$ (Fig.~\ref{fig:fig4}D, right, circles). Next, we disentangle the system again and measure the spin-up probability of Q3 as a function of the evolution time. Without the decoupling pulse, we observe the loss of coherence after a characteristic time $\tau=200$~ns. However, by applying the additional echoing pulse on both Q2 and Q3, we can significantly extend this time scale beyond $2~\mu$s, enough to perform a series of single and multi-qubit gates, owing to our short operation times.

We show this by coherently generating and disentangling a four-qubit Greenberger-Horne-Zeilinger (GHZ) state (Fig.~\ref{fig:fig5}). Making use of the fast two-qubit CZ gates, as well as a decoupling pulse on all qubits, we can maintain phase coherence throughout the experiment. We perform a parity readout on both the Q1Q2 (red) and Q3Q4 (green) at different stages of the algorithm and normalize the observed blocked state fraction to the readout visibility. We prepare a varying initial state by applying a microwave pulse of length $t$ to Q3, as observed in I. After applying CZ gates between all four qubits, the system resides in an entangled GHZ type state at IV/V, for a $\pi/2$ preparation pulse on Q3. The effective spin state oscillates between the antiparallel $\ket{1010}$ and $\ket{0101}$ states as a function of $t_\text{prep}$, resulting in a high state readout for all $t$. The small oscillation that can still be observed for the Q1Q2 system, is caused by a small difference in readout visibility for the two distinct antiparallel spin states. Next, we deploy a Y$^2$ decoupling pulse to echo out all single qubit phase fluctuations during the experiment (Fig.~S11). After disentangling the system again, we project the Q3 qubit state by applying a final X ($\pi/2$) gate, and indeed recover the initial Rabi rotation.

The demonstration of a two-by-two four-qubit array shows that quantum dot qubit systems can be scaled in two-dimensions and multi-qubit logic can be executed. The hole states used are subject to strong spin-orbit coupling, enabling all-electrical driving of the spin state, beneficial for scaling up to even larger systems. Making use of a latched readout mechanism overcomes fast spin relaxation due to the spin-orbit coupling. Furthermore, the ability to freely couple one, two, three and four spins using fast gate pulses has great prospects both for performing high-fidelity quantum gates as well as studying exotic spin systems using analog quantum simulations. While the execution of relevant quantum algorithms will require many more qubits, the germanium platform has the potential to leverage the enormous advancements in semiconductor manufacturing techniques for the realization of fault-tolerant quantum processors.

\section*{Acknowledgements}
We thank L.M.K. Vandersypen for useful discussions and thank S.G.J. Philips for his contributions to software development. M.V. acknowledges support through a Vidi grant, two projectruimte grants, and an NWA grant, all associated with the Netherlands Organization of Scientific Research (NWO).

\section*{Competing Interests}
The authors declare no competing interests. Correspondence should be addressed to M.V. (M.Veldhorst@tudelft.nl).

\section*{Data availability}
All data underlying this study will be available from the 4TU ResearchData repository.

\section*{Supplementary Materials}
\noindent Materials and Methods\\
Tables S1 – S2\\
Figures S1 – S11\\
References (S1 - S14)

\end{document}